# SOME FEATURES OF RAMAN SCATTERING BY MOLECULES ADSORBED ON METAL CRYSTAL FACES AND A FINE LIGHT STRUCTURE

A. M. Polubotko

A.F. Ioffe Physico-Technical Institute Russian Academy of Sciences, Politechnicheskaya 26, 194021, Saint Petersburg, RUSSIA, E-mail: alex.marina@mail.ioffe.ru

## ABSTRACT

The paper analyzes some experiments on Raman scattering by molecules adsorbed on the face (111) of silver monocrystals performed by A. Campion et al. From the existence of the forbidden line $A_{2u}$ of benzene, the conclusion about existence of the surface field, caused by atomic structure of the surface is made. The relatively large intensity of this line allows to make a conclusion about large influence of the electromagnetic field spatial inhomogeneity in crystals on their optical properties. The difference between this field and a regular plane wave, which usually describes propagation of electromagnetic field in solids is named as a fine light structure. The influence of this structure on optical properties of solids is pointed out.

## INTRODUCTION

An electromagnetic field in optics of solid state was considered as a totality of plane waves for a long time. However a new phenomenon, Surface Enhanced Raman scattering (SERS) was discovered in 1974 [1]. Later on another Surface Enhanced processes, such as Surface Enhanced Infrared absorption (SEIRA), Surface Enhanced Resonance Raman scattering (SERRS), Surface Enhanced Hyper Raman Scattering (SEHRS), Enormous SERS or Single Molecule Detection by SERS and some others were found. Their nature was unclear for a long time. However, investigators began to connect these phenomena with the distinction of the electromagnetic field from the fields, represented by plane electromagnetic waves and with excitation of surface plasmons. From our point of view, these phenomena are associated with the existence of surface electromagnetic field, caused by the surface roughness and its enhancement in some regions on the metal surface. In 1987 in [2-6] the conception of a strong light-molecule interaction, which arises in strongly varying surface electromagnetic fields was proposed. This conception allows to explain the enhancement coefficients and many features of SEIRA, SERS, SEHRS and Enormous SERS [3-13]. Another interesting phenomenon is a charge transfer enhancement mechanism, discovered experimentally in [14] on the PMDA molecule (Pyrometallic dianhydride) adsorbed on Cu (111) and Cu (100), which results in enhancement of Raman scattering by molecules adsorbed on single crystal surfaces. The last effect manifests some interesting features. They are face and polarization dependence of Raman band intensities, which was found in [14]. In addition, appearance of forbidden lines of benzene and benzene $d_6$ on single surfaces of Ag (111) and Ag (110) was found in [15, 16] (Figure 1). All these phenomena may be explained by heterogeneous surface electromagnetic fields, arising on single surfaces, which possess by atomic scale roughness. The reason of the enhancement in SEIRA, SERS and SEHRS, is a large increase of the fields and its derivatives $\partial E_{\alpha} / \partial x_{\alpha}$ and also some quantum mechanical features of matrix elements of quadrupole moments with a constant sign [11]. The reason of the enhancement in the charge transfer enhancement mechanism is more complicated and associated both with the change of electron structure and the surface electromagnetic field, caused by surface atomic structure. Some special attention must be made to the results of [15,16]. The authors discuss appearance of forbidden lines, caused by large derivatives $\frac{\partial E_z}{\partial z}$ and due to appearances of sufficiently strong quadrupole interaction. However, they reject this mechanism. From our point of view explanation of the appearance of the forbidden lines of benzene only by the quadrupole interaction is not correct. Since the strong change of the field arises on the atomic scale distances, it is necessary to consider not only the quadrupole interaction, but all the interactions, which are associated with another terms of the expansion of the field into the Taylor series. In general the use of such expansion is not possible and we must consider a precise Hamiltonian of interaction of light with the molecule. The source of the forbidden lines in this case is the total surface field, which is connected with the atomic structure of the surface. As it will be demonstrated below, even application of some simple ideas about this field, results in explanation of appearance of the most strong line, caused by the vibration with the $A_{2u}$ irreducible representation in this system.

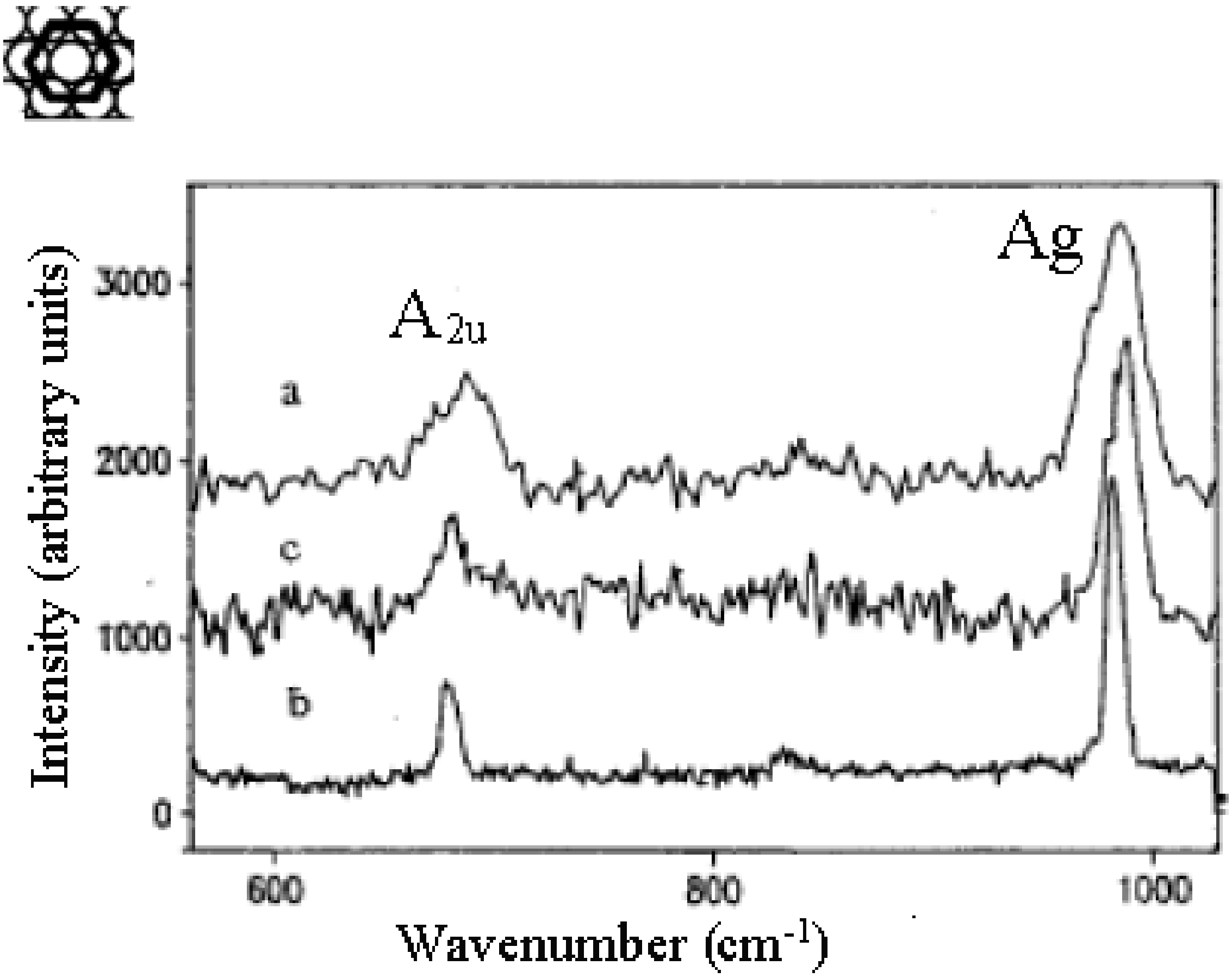

Figure 1. The Raman spectrum of benzene, adsorbed on the face $Ag(111)$ [16] . a) the wavelength of the incident light 633.8 nm, b) 514.5nm, c) 351.1 nm. The geometry of adsorption is shown in the inset.

## RAMAN SCATTERING BY MOLECULES ADSORBED ON FACES OF SILVER SINGLE CRYSTALS AND SURFACE ELECTROMAGNETIC FIELDS CAUSED BY THEIR ATOMIC STRUCTURE

The light-molecule interaction Hamiltonian has the form:

$$-i\frac{e\hbar}{mc}\sum_{i}\overline{A}_{i}\overline{\nabla}_{i}=-i\frac{e\hbar}{mc}\sum_{i}\frac{\overline{E}_{i}}{i\omega}\overline{\nabla}_{i}\ . \tag{1}$$

Here $\overline{A}_i$ is a vector potential, while $\vec{E}_i$ is the electric field at the place of the $i$ electron. There are two fields in a Raman process: the incident $\overline{E}_{inc}$ and scattered $\overline{E}_{scat}$. Here we omit the index $i$, considering that the field is at the place of the $i$ electron. The cross-section of Raman scattering obtained in the framework of the adiabatic perturbation theory can be obtained from the expression, published in [11], using the light-molecule interaction Hamiltonian (1)

$$d\sigma_{s}=\frac{\omega\omega_{s_1}^{3}}{16\hbar^{2}\varepsilon_{0}^{2}\pi^{2}c^{4}}\sum_{p}\left\{\begin{matrix}\dfrac{V_{(s,p)}+1}{2}\\ \dfrac{V_{(s.p)}}{2}\end{matrix}\right\}\left|A_{V_{(s,p)}}\begin{matrix}St\\ anSt\end{matrix}\right|_{vol}^{2}d0 \tag{2}$$

Here

$$A_{V_{(s,p)}}\left\{\begin{matrix}St\\ anSt\end{matrix}\right\}=-\left(\frac{e\hbar}{mc}\right)^{2}\frac{1}{\omega_{scat}\omega_{inc}}\times$$

$$\times[\sum_{\substack{m,l\\ l\neq n}}\frac{R_{n,l,(s,p)}\langle n|\sum_{i}\overline{E}_{i,scat}^{*}\overline{\nabla}_{i}|m\rangle\langle m|\sum_{i}\overline{E}_{i,inc}\overline{\nabla}_{i}|l\rangle}{\left(E_{n}^{(0)}-E_{l}^{(0)}\right)\left(\omega_{mn}\pm\omega_{scat}-\omega_{inc}\right)}+$$

$$+\sum_{\substack{m,l\\ l\neq n}}\frac{R^*_{n,l,(s,p)}\langle l|\sum_i \overline{E}_{i,inc}\overline{\nabla}_i|m\rangle\langle m|\sum_i \overline{E}^*_{i,scat}\overline{\nabla}_i|n\rangle}{\left(E_n^{(0)}-E_l^{(0)}\right)\left(\omega_{mn}\mp\omega_{scat}-\omega_{inc}\right)}+$$

$$+\sum_{\substack{m,l\\ l\neq n}}\frac{R^*_{n,l,(s,p)}\langle l|\sum_i \overline{E}^*_{i,scat}\overline{\nabla}_i|m\rangle\langle m|\sum_i \overline{E}_{i,inc}\overline{\nabla}_i|n\rangle}{\left(E_n^{(0)}-E_l^{(0)}\right)\left(\omega_{mn}-\omega_{inc}\right)}+$$

$$+\sum_{\substack{m,l\\ l\neq n}}\frac{R_{n,l,(s,p)}\langle n|\sum_i \overline{E}_{i,inc}\overline{\nabla}_i|m\rangle\langle m|\sum_i \overline{E}^*_{i,scat}\overline{\nabla}_i|l\rangle}{\left(E_n^{(0)}-E_l^{(0)}\right)\left(\omega_{mn}+\omega_{inc}\right)}+$$

$$\sum_{\substack{m,k\\ m\neq n,k}}\frac{\langle n|\sum_i \overline{E}^*_{i,scat}\overline{\nabla}_i|m\rangle R^*_{m,k,(s,p)}\langle k|\sum_i \vec{E}_{i,inc}\vec{\nabla}_i|n\rangle}{(E_m^{(0)}-E_k^{(0)})(\omega_{mn}\pm\omega_{(s,p)}-\omega_{inc})}+$$

$$\sum_{\substack{m,k\\ m\neq n,k}}\frac{\langle n|\sum_i \overline{E}^*_{i,scat}\overline{\nabla}_i|k\rangle R_{m,k,(s,p)}\langle m|\sum_i \overline{E}_{i,inc}\overline{\nabla}_i|n\rangle}{(E_m^{(0)}-E_k^{(0)})(\omega_{mn}-\omega_{inc})}+$$

$$\sum_{\substack{m,k\\ m\neq n,k}}\frac{\langle n|\sum_i \overline{E}_{i,inc}\overline{\nabla}_i|m\rangle R^*_{m,k,s}\langle k|\sum_i \overline{E}^*_{i,scat}\overline{\nabla}_i|n\rangle}{(E_m^{(0)}-E_k^{(0)})(\omega_{mn}\pm\omega_{(s,p)}+\omega_{scat})}+$$

$$\sum_{\substack{m,k\\ m\neq n,k}}\frac{\langle n|\sum_i \overline{E}_{i,inc}\overline{\nabla}_i|k\rangle R_{mk(s,p)}\langle m|\sum_i \overline{E}^*_{i,scat}\overline{\nabla}_i|n\rangle}{(E_m^{(0)}-E_k^{(0)})(\omega_{mn}+\omega_{scat})}] \qquad (3)$$

Here in (2) and (3) the designations of constants are conventional, $V_{(s,p)}$ is the vibrational quantum number of the $(s,p)$ degenerate vibrational mode. $s$ numerates the degenerate modes, $p$-the modes inside the degenerate mode. $\omega_{inc}$ and $\omega_{scat}$ are the frequencies of the incident and scattered fields, $E_n^{(0)}$ and $E_l^{(0)}$, $l=(l,k,m)$ are the energies of the ground and excited states of the Schrödinger equation for the electron wavefunction with motionless nuclei [3,11]. $\omega_{mn}$ is the difference of the frequencies between $m$ and $n$ states. $R_{n,l,(s,p)}$ and $R_{m,k,(s,p)}$ are the coefficients of excitation of the excited states $l$ and $k$ from the ground state $n$ and the excited state $k$, by the vibrational mode $(s,p)$ [11]. Further it is convenient to introduce a Raman scattering tensor, which has the form

$$C_{V_{(s,p)}}\lfloor f_i, f_j \rfloor = \sum_{\substack{m,l \\ m,l\neq n}} \frac{R_{nl(s,p)}\langle n|f_i|m\rangle\langle m|f_j|l\rangle}{(\mathrm{E}_n^{(0)}-\mathrm{E}_l^{(0)})(\omega_{mn}\pm\omega_{(s,p)}-\omega_{inc})}+$$

$$\sum_{\substack{m,l \\ m,l\neq n}} \frac{R^*_{nl(s,p)}\langle l|f_j|m\rangle\langle m|f_i|n\rangle}{(\mathrm{E}_n^{(0)}-\mathrm{E}_l^{(0)})(\omega_{mn}+\omega_{scat})}+$$

$$\sum_{\substack{m,l \\ m,l\neq n}} \frac{R^*_{nl(s,p)}\langle l|f_i|m\rangle\langle m|f_j|n\rangle}{(\mathrm{E}_n^{(0)}-\mathrm{E}_l^{(0)})(\omega_{mn}-\omega_{inc})}+$$

$$\sum_{\substack{m,l \\ m,l\neq n}} \frac{R_{nl(s,p)}\langle n|f_j|m\rangle\langle m|f_i|l\rangle}{(\mathrm{E}_n^{(0)}-\mathrm{E}_l^{(0)})(\omega_{mn}\pm\omega_{(s,p)}+\omega_{scat})}+$$

$$\sum_{\substack{m,k \\ m\neq n,k}} \frac{\langle n|f_i|m\rangle R^*_{mk(s,p)}\langle k|f_j|n\rangle}{(\mathrm{E}_m^{(0)}-\mathrm{E}_k^{(0)})(\omega_{mn}\pm\omega_{(s,p)}-\omega_{inc})}+$$

$$\sum_{\substack{m,k\\ m\neq n,k}} \frac{\langle n|f_i|k\rangle R_{mk(s,p)}\langle m|f_j|n\rangle}{(\mathrm{E}_m^{(0)}-\mathrm{E}_k^{(0)})(\omega_{mn}-\omega_{inc})}+$$

$$\sum_{\substack{m,k\\ m\neq n,k}} \frac{\langle n|f_j|m\rangle R^*_{mk(s,p)}\langle k|f_i|n\rangle}{(\mathrm{E}_m^{(0)}-\mathrm{E}_k^{(0)})(\omega_{mn}\pm\omega_{(s,p)}+\omega_{scat})}+$$

$$\sum_{\substack{m,k\\ m\neq n,k}} \frac{\langle n|f_j|k\rangle R_{mk(s,p)}\langle m|f_i|n\rangle}{(\mathrm{E}_m^{(0)}-\mathrm{E}_k^{(0)})(\omega_{mn}+\omega_{scat})}+ \quad , \tag{4}$$

where $f_i$ and $f_j$ designate the arguments of the tensor. Its main property, which will be used further is:

$$C_{V_{(s,p)}}[f_i,(a_1 f_j + a_2 f_k)] = a_1 C_{V_{(s,p)}}\lfloor f_i, f_j\rfloor + a_2 C_{V_{(s,p)}}[f_i, f_k] \quad ,$$

$$C_{V_{(s,p)}}[(a_1 f_i + a_2 f_j), f_k] = a_1 C_{V_{(s,p)}}[f_i, f_k] + a_2 C_{V_{(s,p)}}\lfloor f_j, f_k\rfloor \quad .$$

For the Raman scattering by molecules, adsorbed of faces of single crystals, the electromagnetic field consists from the incident, reflected and surface field, caused by atomic scale roughness. The field, which acts on the molecule can be presented as

$$\vec{E}_{inc} = (\vec{E}_{inc})_{vol} + (\vec{E}_{inc})_{surf} + (\vec{E}_{inc,refl})_{vol} \tag{5}$$

and the field, scattered by the molecule, as

$$\vec{E}_{scat} = (\vec{E}_{scat})_{vol} + (\vec{E}_{scat})_{surf} + (\vec{E}_{scat,refl})_{vol} \tag{6}$$

The corresponding expressions have the form

$$(\vec{E}_{inc})_{vol} = \vec{e}_{o,inc} e^{ik_{0x}x + ik_{0y}y - ik_{0z}z} \quad , \tag{7}$$

$$(\vec{E}_{inc})_{surf} = e^{i\vec{k}_0\vec{r}} \sum_{\vec{\bar{k}}} \vec{a}_{inc,\bar{k}} e^{i\bar{k}\vec{r} + i\sqrt{k_0^2 - \bar{k}^2}\, z} \quad , \tag{8}$$

$$(\vec{E}_{inc,refl})_{vol} = \vec{e}_{scat,refl} e^{ik_{0x}x + ik_{0y}y + ik_{0z}z} \quad . \tag{9}$$

Here the sign $vol$ in (7) means that we deal with a plane homogeneous wave, which propagates in a free space. The same refers to the expression (9). The sign $surf$ in (8) designates that we deal with surface electromagnetic field. Analogous expressions can be written out for the field, which is scattered by the molecule

$$(\vec{E}_{scat})_{vol} = \vec{e}_{o,scat} e^{ik_{0x}x + ik_{0y}y - ik_{0z}z}$$

$$(\vec{E}_{scat})_{surf} = e^{i\vec{k}_0\vec{r}} \sum_{\vec{k}} \vec{a}_{scat,\bar{k}} e^{i\vec{k}\vec{r} + i\sqrt{k_0^2 - \bar{k}^2}\, z} \qquad (10)$$

$$(\vec{E}_{scat,refl})_{vol} = \vec{e}_{scat,refl} e^{ik_{0x}x + ik_{0y}y + ik_{0z}z}$$

Neglecting by the factor $e^{i\bar{k}_0\bar{r}}$, formulae (8) and (10) represents expansion of the surface electromagnetic field in space harmonics with the amplitudes $\bar{a}_k$ and the wave vectors $\bar{k} = (k_x = \frac{2\pi n_1}{a_1}, k_y = \frac{2\pi n_2}{a_2})$, where $n_1, n_2$ are integers, which vary in the interval $(-\infty, +\infty)$, $a_1, a_2$ are the dimensions of an elementary sell in the $x$ and $y$ directions. One should note that after the neglect by the dependence on the incident field, the components of this field transform after irreducible representations of the surface symmetry group and are localized approximately in a very narrow region $l/2\pi$, where $l = \sqrt{(a_1^2 + a_2^2)}$ is the characteristic size of the surface cell. It is obvious, that the light-molecule interaction Hamiltonian depends on polarization of the incident field $(\bar{E}_{inc})_{vol}$ and on the geometry of a crystal face via $(\bar{E}_{inc})_{surf}$. However apparently the strong localization of $(\bar{E}_{inc})_{surf}$ in the narrow surface region results only in a slight change of separate Raman bands with respect to those received in usual Raman scattering. Experiments, which prove the existence of the surface field, caused by the surface atomic roughness were performed in [15, 16]. The authors observed the lines of benzene and benzene $d_6$, adsorbed on silver, which refer to the irreducible representation $A_{2u}$, and are forbidden in usual Raman scattering (Figure 1). In order to reveal the reason of appearance of these lines, it is necessary to consider the symmetry properties of the light-molecule interaction Hamiltonian for the incident and scattered fields (1) in detail, taking into account their properties (5,6). First of all it is necessary to note that the tangential components of the electric field are small because of the screening by conductivity electrons of metal. This condition must be fulfilled on the surface sufficiently well, while the normal component $E_z$ is sufficiently strong, especially for the surface field, where its increase is associated with a large local curvature on the atomic structure of the surface. Such values as $(\vec{E}_{inc})_{vol} + (\vec{E}_{inc,refl})_{vol}$ and $(\vec{E}_{scat})_{vol} + (\vec{E}_{scat,refl})_{vol}$ have the characteristic size $\sim\lambda$, or of the same order of magnitude as the wavelength of the incident field that is significantly larger and is not comparable with the dimensions of the elementary surface cell, while the characteristic size of $(E_{inc})_{surf}$ and $(E_{scat})_{surf}$ is $l$, or the size of the elementary cell. These results allow to use a dipole approximation of the interaction Hamiltonian for two first two types of the fields, while we must use precise field structure of the surface field in the elementary cell because its strong change. The symmetry of the $E_z$ component coincides with the symmetry of the surface cell, neglecting by the change, associated with the slow changing factor $e^{i\vec{k}_0\vec{r}}$. The light-molecule interaction Hamiltonian (for the incident field for example) has the form

$$\widehat{\mathbf{H}}_{mol-r} = \frac{(d_z(E_{z,inc})_{vol})e^{-i\omega_{inc}t} + (d_z(E^*_{z,inc})_{vol})e^{i\omega_{inc}t}}{2}$$

$$-i\frac{e\hbar}{m}\sum_i \frac{(E_{i,z,inc})_{surf}\, e^{-i\omega_{inc}t} + (E^*_{i,z,inc})_{surf}\, e^{i\omega_{inc}t}}{2i\omega_{inc}} \nabla_{i,z} \qquad (11)$$

Then, taking into account the previous consideration and (11), the expression for $A_{V_{(s,p)}}\left\{\begin{matrix} St \\ anSt \end{matrix}\right\}$ in (2) can be presented by several contributions

$$A_{V_{(s,p)}}\left\{\begin{matrix} St \\ anSt \end{matrix}\right\} = C_{V_{(s,p)}}[d_z, d_z] \times (E_{z,inc})_{vol} (E^*_{z,scat})_{vol} +$$

$$-\frac{e\hbar}{m\omega_{scat}} C_{V_{(s,p)}}[d_z, \sum_i ((E^*_{i,z,scat})_{surf} \nabla_{i,z})] \times (E_{z,inc})_{vol} +$$

$$-\frac{e\hbar}{m\omega_{inc}} C_{V_{(s,p)}}[\sum_i ((E_{i,z,inc})_{surf} \nabla_{i,z}), d_z] \times (E^*_{z,scat})_{vol} +$$

$$+\left(\frac{e\hbar}{m}\right)^2 \frac{1}{\omega_{inc}\omega_{scat}} C_{V_{(s,p)}}[\sum_i ((E_{i,z,inc})_{surf} \nabla_{i,z}), \sum_i ((E^*_{i,z,scat})_{surf} \nabla_{i,z}]$$

(12)

The first contribution is caused by absorption of the regular plane wave and by emission of the regular plane wave also. The second one is the absorption of the plane wave and emission of the surface field, the third contribution is absorption of the surface field and emission of the plane scattered wave. The forth contribution is absorption and emission of pure surface fields.

Further we shall investigate the Raman spectrum of benzene and benzene $d_6$, adsorbed of the face of $Ag$ (111). Selection rules for the scattering tensor (4) can be obtained from the following expression for one of the lines in (4)

$$R_{n.l.(s,p)}\langle n|f_i|m\rangle\langle m|f_j|l\rangle \neq 0$$

which results in the following conditions

$$R_{n.l.(s,p)} \neq 0 \quad (13a)$$

$$\langle n|f_i|m\rangle \neq 0 \quad (13b)$$

$$\langle m|f_j|l\rangle \neq 0 \quad (13c)$$

and must be satisfied simultaneously. The condition (13a) determines an irreducible representation, which describes the vibrational mode $(s,p)$ [6,11]

$$\Gamma_{(s,p)} \in \Gamma_l \Gamma_n \quad . \tag{14}$$

Here the sign $\Gamma$ designates irreducible representations, which describe transformational properties of corresponding values, the vibrational mode $(s,p)$ and the states $l$ and $n$. The condition (13c) determines the irreducible representation of the state $l$ [17]

$$\Gamma_l \in \Gamma_{f_j} \Gamma_m \quad , \tag{15}$$

and the condition (13b) determines the irreducible representation of the state $m$

$$\Gamma_m \in \Gamma_n \Gamma_{f_i} \quad . \tag{16}$$

Substituting (16) in (15) and further in (14) one can obtain selection rule for the tensor

$$\Gamma_{(s,p)} \in \Gamma_{f_i} \Gamma_{f_j} \quad .$$

The most probable adsorption geometry is shown on Figure 1 (the upper left corner). The surface fields $(E_{i,z,inc})_{surf}$ and $(E_{i,z,scat})_{surf}$ have the symmetry, which is described by the symmetry group $C_6$. The value $\sum_i ((E_{i,z,inc})_{surf} \nabla_{i,z})$ can be expressed as

$$\sum_i ((E_{i,z,inc})_{surf} \nabla_{i,z}) = A^+_{inc} + A^-_{inc} \ , \tag{18}$$

where

$$A^+_{inc} = \frac{1}{2}\left(\sum_i ((E_{i,z,inc})_{surf} \nabla_{i,z}) + \sum_i ((E_{i,-z,inc})_{surf} \nabla_{i,-z})\right) \tag{19}$$

$$A^-_{inc} = \frac{1}{2}\left(\sum_i ((E_{i,z,inc})_{surf} \nabla_{i,z}) - \sum_i ((E_{i,-z,inc})_{surf} \nabla_{i,-z})\right) \ . \tag{20}$$

Similar expressions can be written for the scattered fields. Then one can see, that the value $A^+_{inc}$ (19) transforms after the unit irreducible representation of the symmetry group of benzene $D_{6h}$, while $A^{-}_{inc}$ (20) transforms after the irreducible representation $A_{2u}$. Substituting (18) and analogous expression for the scattered field in (12), using the selection rule for the tensor (17), one can see that (12) contributes in $S_{A_g}$ in the lines, caused by vibrations with the unit irreducible representation and also the contributions $S_{A_{2u}}$, caused by vibrations with the irreducible representation $A_{2u}$.

$$A_{V_{(s,p)}}\begin{Bmatrix} St \\ anSt \end{Bmatrix} = S_{A_g} + S_{A_{2u}} ,$$

where

$$S_{A_g} = C_{V_{(s,p)}}[d_z, d_z] \times (E_{z,inc})_{vol}(E^*_{z,scat})_{vol} +$$

$$-\frac{e\hbar}{m\omega_{scat}} C_{V_{(s,p)}}[d_z, A^{-*}_{scat})] \times (E_{z,inc})_{vol} +$$

$$-\frac{e\hbar}{m\omega_{inc}} C_{V_{(s,p)}}[A^{-}_{inc}, d_z] \times (E^*_{z,scat})_{vol} +$$

$$+\left(\frac{e\hbar}{m}\right)^2 \frac{1}{\omega_{inc}\omega_{scat}} C_{V_{(s,p)}}[A^{-}_{inc}, A^{-*}_{scat}] -$$

$$+\left(\frac{e\hbar}{m}\right)^2 \frac{1}{\omega_{inc}\omega_{scat}} C_{V_{(s,p)}}[A^{+}_{inc}, A^{+*}_{scat}] ,$$

$$S_{A_{2u}} = -\frac{e\hbar}{m\omega_{scat}} C_{V_{(s,p)}}[d_z, A^{+*}_{scat})] \times (E_{z,inc})_{vol} +$$

$$-\frac{e\hbar}{m\omega_{inc}} C_{V_{(s,p)}}[A^{+}_{inc}, d_z] \times (E^*_{z,scat})_{vol} +$$

$$+\left(\frac{e\hbar}{m}\right)^2 \frac{1}{\omega_{inc}\omega_{scat}} C_{V_{(s,p)}}[A^{+}_{inc}, A^{-*}_{scat}] +$$

$$+\left(\frac{e\hbar}{m}\right)^2 \frac{1}{\omega_{inc}\omega_{scat}} C_{V_{(s,p)}} [A^{-}_{inc}, A^{+*}_{scat}] \ .$$

In case, when the surface field is absent, the expression (12) would contain only the first term with the value $C_{V_{(s,p)}}[d_z, d_z]$, which is not equal to zero only for the totally symmetric vibrations, transforming after the unit irreducible representation. Thus, appearance of the benzene line with the irreducible representation $A_{2u}$, which is forbidden in the usual Raman scattering , is associated with the existence of the surface field, caused by the atomic surface roughness. One should note that the surface field influence mainly on the chemisorbed molecules, when the molecule penetrates strongly in the surface region. Most previous results of A. Campion, who discovered the absence of the enhancement of Raman scattering on a large number of molecules apparently refer mainly to the physisorbed molecules, which do not penetrate strongly in the region of the surface field. A sharp difference from these works is observed in [14], where the PMDA molecule is considered. Strong chemisorption of the PMDA molecule, which results in the change of its geometry and strong lowering of its first electron level apparently is the reason of the resonance enhancement and manifestation of the enhancement mechanism, associated with charge transfer.

The strong change of the surface electromagnetic field in the above mentioned processes , especially in the enhancement of the process with the charge transfer and its strong difference from the field in a free space can be named as a fine light structure. This structure exists in solids because of the inhomogeneity of the atomic structure, when the electromagnetic field can not be presented by a regular plane wave in the substance. Existence of the forbidden band of benzene $A_{2u}$, adsorbed on the face $Ag$ (111) and its sufficiently strong intensity, compared with the line, caused by the breathing mode [16], allows us to make a conclusion about essential effect of the fine light structure on the optical properties of solids with a large values of dielectric constants, where the surface fields must be sufficiently strong. It is necessary to note that this fact can be corroborated by the difference of optical constants of various dielectrics, which consist from various atoms and have various inhomogeneous structures. Thus, this structure determines the optical properties in a large number of optical phenomena and can be essential in optics.